\documentclass[runningheads]{llncs}
\usepackage[T1]{fontenc}

\usepackage{graphicx}
\usepackage{tikz}
\usepackage{float}
\usepackage{amsmath}

\begin{document}
\title{Network Design on Undirected Series-Parallel Graphs}
%
%\titlerunning{Network Design on SP Graphs}

\author{Ishan Bansal\inst{1}\orcidID{0000-0002-5083-309X} \and
Ryan Mao\inst{1}\orcidID{0009-0006-6288-1052} \and
Avhan Misra\inst{1}\orcidID{0009-0004-6657-5947}}
\authorrunning{I. Bansal et al.}
\institute{Cornell University, NY, USA \email{\{ib332,rwm275,am2934\}@cornell.edu}}
\maketitle

\begin{abstract}
We study the single pair capacitated network design problem and the budget constrained max flow problem on undirected series-parallel graphs. These problems were well studied on directed series-parallel graphs, but little is known in the context of undirected graphs. The major difference between the cases is that the source and sink of the problem instance do not necessarily coincide with the terminals of the underlying series-parallel graph in the undirected case, thus creating certain complications. We provide pseudopolynomial time algorithms to solve both of the problems and provide an FPTAS for the budget constrained max flow problem. We also provide some extensions, arguing important cases when the problems are polynomial-time solvable, and describing a series-parallel gadget that captures an edge upgrade version of the problems.

\end{abstract}

\section{Introduction}

Network design problems have played a central role in the field of combinatorial optimization and naturally arise in various applications, including transportation, security, communications, and supply-chain logistics. In this paper, we study two problems in the area of network design, the \textit{Budget-Constrained Max Flow Problem} (BCMFP) and the single pair \textit{Capacitated Network Design Problem} (CapNDP). In both these problems, the input is a multi-graph with costs and capacities on edges, and a source-sink pair of vertices. The algorithmic goal in the BCMFP is to maximize the connectivity between the source and sink while the total cost of purchased edges fits within a given budget. CapNDP is the minimization version of BCMFP where the goal is to find a cheapest subgraph that meets a demanded level of connectivity between the source and sink.

We study the version of these problems where the costs on edges are all-or-nothing. Hence, one has to pay for the entire edge, irrespective of the amount of positive flow being sent across it. These problems are notoriously difficult, and we limit our focus to a class of input graphs that emerge as specific types of electrical networks, known as series-parallel graphs, which possess favorable algorithmic properties. Both the problems BCMFP and CapNDP have been well studied on directed series-parallel graphs \cite{krumke1999flow,schwarz1998budget} with fully polynomial time approximation schemes (FPTAS) available. However, the situation is more complex in the case of undirected series-parallel graphs as noted in \cite{carr1999strengthening}. We present a pseudopolynomial time algorithm for both BCMFP and CapNDP on undirected series parallel graphs and present an FPTAS for BCMFP. We supplement these results with some extensions of these problems, and show that our pseudo-polynomial time algorithm can be implemented in polynomial time if the input capacities are suitably structured.  Note that both of the problems are \textsc{NP-Hard} even on simple two vertex graphs, since they capture the knapsack and the min-knapsack problems. 

\subsection{Related Work}

The budget constrained max flow problem was studied by Schwarz and Krumke \cite{schwarz1998budget}. They considered different cost variants of the problem including the most difficult all-or-nothing variant. Here, they provided a pseudopolynomial time algorithm when the input graph is a directed series-parallel graph and used common scaling techniques to further provide an FPTAS. They also show that the problem is \textsc{NP-Hard} on series parallel graphs and strongly \textsc{NP-Hard} even on bipartite graphs. Krumke et al. \cite{krumke1999flow} obtained similar results for the single pair capacitated network design problem on directed series parallel graphs. 

The main structural tool used in these results was developed by Valdes et al. \cite{valdes1979recognition}, where they provide a linear time algorithm to produce a decomposition of a series-parallel graphs. Furthermore, by preprocessing the input graph, one can always assume that the terminals of the series parallel graph coincide with the source and sink of the problem instance \cite{schwarz1998budget}. Carr et al. \cite{carr1999strengthening} observed that the case is slightly more complex for undirected series-parallel graphs as now the terminals of the graph and the source-sink pair of the problem instance need not coincide. They provided a pseudopolynomial time algorithm for the CapNDP problem on outerplanar graphs (a sub-family of series-parallel graphs), but incorrectly claimed that this implies an FPTAS for CapNDP on outerplanar graphs. We outline this difficulty in Section 4. 

Carr et al. \cite{carr1999strengthening} also provided an approximation algorithm for the general CapNDP problem by strengthening the base LP using the so called knapsack cover constraints. However, no $o(m)$-approximation algorithm is known for the general CapNDP problem where $m$ is the number of edges in the input graph. Chakraborty et al. \cite{chakrabarty2015approximability} showed that CapNDP cannot be approximated to within $\Omega(\log\log m)$ unless $NP\subseteq DTIME(m^{\log\log\log m})$.

In recent years, a generalization of BCMFP called budget constrained minimum cost flows has received attention \cite{holzhauser2016budget,holzhauser2017complexity,holzhauser2017network}. Here, the algorithmic goal is to compute a classical minimum cost-flow while also adhering to a flow-usage budget.

\subsection{Our Results}

We provide a pseudopolynomial time algorithm for BCMFP on undirected series parallel graphs in Section 3 by designing a dynamic program that makes use of the series-parallel decomposition of a series-parallel graph. In the case of directed series-parallel graphs, \cite{schwarz1998budget} considered the minimum cost to flow a value of $f$ between the terminals of a series-parallel graph, and computed this minimum cost using the decomposition of the series-parallel graph. However as also noted in \cite{carr1999strengthening}, the terminals of an undirected series parallel graph need not coincide with the source and the sink of the problem instance, and so the method used in \cite{schwarz1998budget} does not work. Instead, we make use of circulations, and consider the minimum cost to flow a suitably defined circulation in the series-parallel graph. We show how all arising cases can still be handled using a more complicated dynamic program.  Since the CapNDP problem is just the minimization version of the BCMFP, this also implies that CapNDP on series parallel graphs can be solved in pseudopolynomial time.

\begin{theorem}\label{thm:pseudo}
    There exists pseudopolynomial time algorithms to solve both the BCMFP and the single pair CapNDP on undirected series-parallel graphs.
\end{theorem}

In Section 4, we leverage the above algorithm to provide an FPTAS for the BCMFP on series-parallel graphs. This is done using common scaling techniques that lowers the capacities on edges so that their values are polynomially bounded. The pseudopolynomial time algorithm then runs in polynomial time as long as all capacities are still integral. But in maintaining integrality, we must lose an $\epsilon$ factor in optimality. However, in contrast to the results in \cite{krumke1999flow}, a similar FPTAS cannot be obtained for the CapNDP. This is rather intriguing and pinpoints a difficulty in extending earlier results on directed series-parallel graphs to undirected series-parallel graphs.

\begin{theorem}\label{thm:fptas}
    There exists a fully polynomial time approximation scheme for the BCMFP on series-parallel graphs
\end{theorem}

In section 5, we consider two extensions of our results. First, we provide instances when the pseudopolynomial algorithm developed in section 3 can be implemented in polynomial time. For instance, if all the capacities on edges are the same, then we can scale down the capacities so that they are all unit (and hence polynomially bounded) without loosing any factor in optimality. This cannot be done even if there are just two distinct values of capacities. Nonetheless, we show that our algorithm can still be implemented in polynomial time in such and more general cases.

\begin{theorem}\label{thm:lattice}
    Suppose the capacities on edges lie in a bounded lattice defined by basis values $d_1,\ldots,d_k$ i.e. $\{\sum_{i=1}^k \alpha_id_i : |\alpha_i|\leq K\}$ where $k$ is a constant and $K$ is polynomially bounded, then both the problems BCMFP and CapNDP can be solved in polynomial time.
\end{theorem}

Second, we consider a variant of the BCMFP and CapNDP where edges have various levels of upgrades with different cost and capacity values, but only one upgrade can be chosen and purchased. Using parallel edges for each upgrade does not solve this problem since multiple upgrades could then be simultaneously applied. Instead, we provide a simple gadget that is series-parallel and that captures this extension. Hence, all our earlier results hold in this setting as well.

\section{Preliminaries}

Now some basic definitions and problem statements.
\begin{definition}[Graph]
    A graph $G = (V,E)$ is a set of vertices $V$ and possibly parallel edges $E$ with non-negative costs $\{c_e\}$ and non-negative capacities $\{u_e\}$ on the edges. Additionally, there are two designated vertices $s$ and $t$ called the source and the sink. \\
    The cost of a subset of the edges $F\subseteq E$ denoted $c(F)$ is defined as $\sum_{e\in F}c_e$. We denote the number of edges $|E|$ using $m$ and the number of vertices $|V|$ using $n$.
\end{definition}

\begin{definition}[Flow]
    An $s$-$t$ flow in a graph $G$ is an assignment of a direction and a non-negative flow value $f_e \leq u_e$ on each edge such that for every vertex $v$ other than $s$ and $t$, the sum of the flow values on edges directed into $v$ is equal to the sum of the flow values on edges directed out of $v$. The value of the $s$-$t$ flow is equal to the sum of the flow values on edges directed out of $s$.
\end{definition}

\begin{definition}[Circulation]
    Given a graph $G$ and a set of values $\{r_v\}$ called residues on the vertices, an $r$-circulation is an assignment of a direction and a non-negative flow value $f_e \leq u_e$ on each edge such that for every vertex $v$, the sum of the flow values on edges directed into $v$ minus the sum of the flow values on edges directed out of $v$ is equal to $r_v$.
\end{definition}

\begin{definition}[BCMFP]
    An instance of the budget constrained max flow problem (BCMFP) takes as input a graph $G = (V,E)$ and a budget $B$. The goal is to find a set of edges $F$ such that $c(F) \leq B$ that maximizes the max $s$-$t$ flow in the graph $(V,F)$.
\end{definition}

\begin{definition}[CapNDP]
    An instance of the capacitated network design problem (CapNDP) takes as input a graph $G = (V,E)$ and a demand $D$. The goal is to find a cheapest set of edges $F$, such that the max $s$-$t$ flow in the graph $(V,F)$ is at least the demand $D$.
\end{definition}

\begin{definition}[Series-Parallel Graphs]
    A series-parallel graph is defined recursively as follows. The graph $G$ with vertex set $\{a,b\}$ and edge set $\{(a,b)\}$ is a series-parallel graph with terminals $a$ and $b$. If $G_1=(V_1,E_1)$ and $G_2=(V_2,E_2)$ are series-parallel with terminals $a_1,b_1$ and $a_2,b_2$ respectively, then
\begin{itemize}
    \item The graph obtained by identifying $b_1$ and $a_2$ is series-parallel with terminals $a_1$ and $b_2$. This graph is the series composition of $G_1$ and $G_2$. 
    \item The graph obtained by identifying $a_1$ with $a_2$ and also $b_1$ with $b_2$ is series-parallel with terminals $a_1(=a_2)$ and $b_1(=b_2)$. This graph is the parallel composition of $G_1$ and $G_2$.
\end{itemize}
\end{definition}

There are well known algorithms to determine if a graph is series-parallel, and if so, provide a parse tree (decomposition tree) specifying how the graph is obtained using the above two composition rules \cite{valdes1979recognition}.

\begin{definition}[FPTAS]
    A fully polynomial time approximation scheme (FPTAS) for a maximization problem is an algorithm which takes as input a problem instance and a parameter $\epsilon>0$ and returns a solution which is within a $1-\epsilon$ factor of the optimal solution, and runs in time polynomial in the size of the instance and $1/\epsilon$.
\end{definition}

\section{Pseudopolynomial Algorithm}

In this section we present a pseudopolynomial time algorithm to solve BCMFP and CapNDP on undirected series-parallel graphs, thus proving Theorem \ref{thm:pseudo}.

\begin{proof}
Let $F$ be the maximum possible $s$-$t$ flow in the graph $G$. We will compute suitably defined circulations in subgraphs of $G$ that arise in its series-parallel decomposition. Let $G'=(V',E')$ be a subgraph of $G$ which is in the series-parallel decomposition tree of $G$. Let $a$ and $b$ be its terminals. We associate with $G'$ a 2, 3 or 4-tuple $R_{G'}=(r_{a},r_s,r_t,r_{b})$ where the term $r_s$ is included only if $G'$ contains $s$ and $s\neq a,b$, and similarly for the the term $r_t$. The states of our dynamic program (DP) will be $(G',R_{G'})$ for all possible tuples $R_{G'}$ such that $r\in R_{G'}\Rightarrow |r|\leq F$ and $\sum_{r\in R_{G'}}r=0$. Note that the tuple $R_{G'}$ also defines the residues of a circulation in $G'$ by setting $r_v = 0$ for all $v\in G \setminus \{a,s,t,b\}$.

For each such state, we find the minimum cost subset of edges $F'\subseteq E'$ such that the subgraph $(V',F')$ admits an $R_{G'}$-circulation. When it is clear from context, we abuse notation to denote both the set of edges $F'$ and its cost $c(F')$ as $DP(G',R_{G'})$. We now use the series-parallel decomposition tree of $G$ to find $DP(G',R_{G'})$. Clearly if $G'$ was a leaf of the decomposition tree, then we could find $F'$ very easily (as every leaf of the decomposition tree is a two vertex graph with one edge). We consider the following cases.
\begin{itemize}
    \item \textbf{Case 1} $R_{G'}$ is a 2-tuple i.e $R_{G'}=(r_{a},r_{b})$ \\
    \begin{itemize}
        \item \textbf{Case 1.1} $G'$ is formed using a series composition of $G_1$ and $G_2$.
        
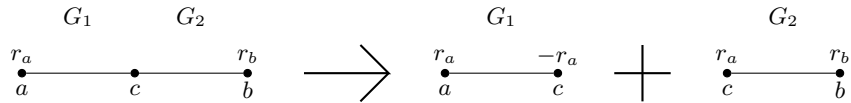
\begin{figure}[H]
            \centering
            \begin{tikzpicture}[scale = 0.75]
                \node[circle, draw, fill = black, inner sep=1pt] at (0,0) (node a) {}; \node at (0,-0.3) {\small $a$}; \node at (0,0.3) {\small $r_{a}$};
                \node[circle, draw, fill = black, inner sep=1pt] at (2,0) (node c) {}; \node at (2,-0.3) {\small $c$};
                \node[circle, draw, fill = black, inner sep=1pt] at (4,0) (node b) {}; \node at (4,-0.3) {\small $b$}; \node at (4,0.3) {\small $r_{b}$};
                \draw[thin] (node a) -- (node c) -- (node b);
                \node at (1,1) {$G_1$}; \node at (3,1) {$G_2$};
                \draw[thick] (5,0) -- (6.5,0) -- (6,0.5);\draw[thick] (6.5,0) -- (6,-0.5);
                \node[circle, draw, fill = black, inner sep=1pt] at (7.5,0) (node a2) {}; \node at (7.5,-0.3) {\small $a$}; \node at (7.5,0.3) {\small $r_{a}$};
                \node[circle, draw, fill = black, inner sep=1pt] at (9.5,0) (node c2) {}; \node at (9.5,-0.3) {\small $c$}; \node at (9.5,0.3) {\small $-r_{a}$};
                \node at (8.5,1) {$G_1$}; \draw[thin] (node a2) -- (node c2);
                \draw[thick] (10.5,0) -- (11.5,0);\draw[thick] (11,0.5) -- (11,-0.5);
                \node[circle, draw, fill = black, inner sep=1pt] at (12.5,0) (node c3) {}; \node at (12.5,-0.3) {\small $c$}; \node at (12.5,0.3) {\small $r_{a}$};
                \node[circle, draw, fill = black, inner sep=1pt] at (14.5,0) (node b3) {}; \node at (14.5,-0.3) {\small $b$}; \node at (14.5,0.3) {\small $r_{b}$};
                \node at (13.5,1) {$G_2$}; \draw[thin] (node c3) -- (node b3);
            \end{tikzpicture}
        \end{figure}
        \vspace{-5mm}
        $$DP(G',R_{G'}) = DP(G_1, (r_{a},-r_{a})) + DP(G_2, (r_{a},r_{b}))$$ \\

        \item \textbf{Case 1.2} $G'$ is formed using a parallel composition of $G_1$ and $G_2$.
        
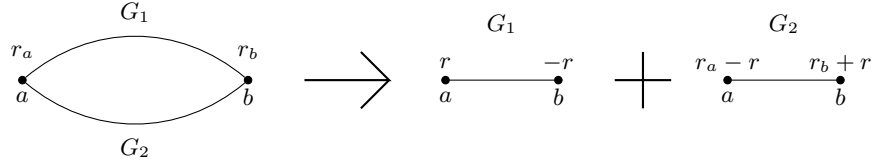
\begin{figure}[H]
            \centering
            \begin{tikzpicture}[scale = 0.75]
                \node[circle, draw, fill = black, inner sep=1pt] at (0,0) (node a) {}; \node at (0,-0.3) {\small $a$}; \node at (0,0.5) {\small $r_{a}$};
                \node[circle, draw, fill = black, inner sep=1pt] at (4,0) (node b) {}; \node at (4,-0.3) {\small $b$}; \node at (4,0.5) {\small $r_{b}$};
                \draw[bend right = 40,thin] (node a) to (node b); \draw[bend left = 40,thin] (node a) to (node b);
                \node at (2,1.2) {$G_1$}; \node at (2,-1.2) {$G_2$};
                \draw[thick] (5,0) -- (6.5,0) -- (6,0.5);\draw[thick] (6.5,0) -- (6,-0.5);
                \node[circle, draw, fill = black, inner sep=1pt] at (7.5,0) (node a2) {}; \node at (7.5,-0.3) {\small $a$}; \node at (7.5,0.3) {\small $r$};
                \node[circle, draw, fill = black, inner sep=1pt] at (9.5,0) (node c2) {}; \node at (9.5,-0.3) {\small $b$}; \node at (9.5,0.3) {\small $-r$};
                \node at (8.5,1) {$G_1$}; \draw[thin] (node a2) -- (node c2);
                \draw[thick] (10.5,0) -- (11.5,0);\draw[thick] (11,0.5) -- (11,-0.5);
                \node[circle, draw, fill = black, inner sep=1pt] at (12.5,0) (node c3) {}; \node at (12.5,-0.3) {\small $a$}; \node at (12.5,0.3) {\small $r_{a}-r$};
                \node[circle, draw, fill = black, inner sep=1pt] at (14.5,0) (node b3) {}; \node at (14.5,-0.3) {\small $b$}; \node at (14.5,0.3) {\small $r_{b} + r$};
                \node at (13.5,1) {$G_2$}; \draw[thin] (node c3) -- (node b3);
            \end{tikzpicture}
        \end{figure}
        \vspace{-10mm}
        $$ DP(G',R_{G'}) = \min_{-F\leq r\leq F} DP(G_1,(r,-r)) + DP(G_2, (r_{a} - r, r_{b}+r))$$
    \end{itemize} ~\\

    \item \textbf{Case 2} $R_{G'}$ is a 3-tuple i.e $R_{G'}=(r_{a},r_s,r_{b})$ \\
    
    \begin{itemize}
        \item \textbf{Case 2.1} $G'$ is formed using a series composition of $G_1$ and $G_2$ splitting at $s$.
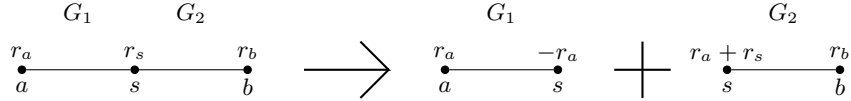
        \begin{figure}[H]
            \centering
            \begin{tikzpicture}[scale = 0.75]
                \node[circle, draw, fill = black, inner sep=1pt] at (0,0) (node a) {}; \node at (0,-0.3) {\small $a$}; \node at (0,0.3) {\small $r_{a}$};
                \node[circle, draw, fill = black, inner sep=1pt] at (2,0) (node c) {}; \node at (2,-0.3) {\small $s$}; \node at (2,0.3) {\small $r_{s}$};
                \node[circle, draw, fill = black, inner sep=1pt] at (4,0) (node b) {}; \node at (4,-0.3) {\small $b$}; \node at (4,0.3) {\small $r_{b}$};
                \draw[thin] (node a) -- (node c) -- (node b);
                \node at (1,1) {$G_1$}; \node at (3,1) {$G_2$};
                \draw[thick] (5,0) -- (6.5,0) -- (6,0.5);\draw[thick] (6.5,0) -- (6,-0.5);
                \node[circle, draw, fill = black, inner sep=1pt] at (7.5,0) (node a2) {}; \node at (7.5,-0.3) {\small $a$}; \node at (7.5,0.3) {\small $r_{a}$};
                \node[circle, draw, fill = black, inner sep=1pt] at (9.5,0) (node c2) {}; \node at (9.5,-0.3) {\small $s$}; \node at (9.5,0.3) {\small $-r_{a}$};
                \node at (8.5,1) {$G_1$}; \draw[thin] (node a2) -- (node c2);
                \draw[thick] (10.5,0) -- (11.5,0);\draw[thick] (11,0.5) -- (11,-0.5);
                \node[circle, draw, fill = black, inner sep=1pt] at (12.5,0) (node c3) {}; \node at (12.5,-0.3) {\small $s$}; \node at (12.5,0.3) {\small $r_{a}+ r_s$};
                \node[circle, draw, fill = black, inner sep=1pt] at (14.5,0) (node b3) {}; \node at (14.5,-0.3) {\small $b$}; \node at (14.5,0.3) {\small $r_{b}$};
                \node at (13.5,1) {$G_2$}; \draw[thin] (node c3) -- (node b3);
            \end{tikzpicture}
        \end{figure}
        \vspace{-5mm}
        $$DP(G',R_{G'}) = DP(G_1,(r_{a},-r_{a})) + DP(G_2, (r_{a}+r_s,r_{b}))$$ \\
        \item \textbf{Case 2.2} $G'$ is formed using a series composition of $G_1$ and $G_2$ not splitting at $s$.
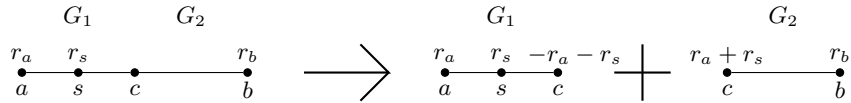
        \begin{figure}[H]
            \centering
            \begin{tikzpicture}[scale = 0.75]
                \node[circle, draw, fill = black, inner sep=1pt] at (0,0) (node a) {}; \node at (0,-0.3) {\small $a$}; \node at (0,0.3) {\small $r_{a}$};
                \node[circle, draw, fill = black, inner sep=1pt] at (1,0) (node s) {}; \node at (1,-0.3) {\small $s$}; \node at (1,0.3) {\small $r_{s}$};
                \node[circle, draw, fill = black, inner sep=1pt] at (2,0) (node c) {}; \node at (2,-0.3) {\small $c$};
                \node[circle, draw, fill = black, inner sep=1pt] at (4,0) (node b) {}; \node at (4,-0.3) {\small $b$}; \node at (4,0.3) {\small $r_{b}$};
                \draw[thin] (node a) -- (node c) -- (node b);
                \node at (1,1) {$G_1$}; \node at (3,1) {$G_2$};
                \draw[thick] (5,0) -- (6.5,0) -- (6,0.5);\draw[thick] (6.5,0) -- (6,-0.5);
                \node[circle, draw, fill = black, inner sep=1pt] at (7.5,0) (node a2) {}; \node at (7.5,-0.3) {\small $a$}; \node at (7.5,0.3) {\small $r_{a}$};
                \node[circle, draw, fill = black, inner sep=1pt] at (8.5,0) (node s2) {}; \node at (8.5,-0.3) {\small $s$}; \node at (8.5,0.3) {\small $r_{s}$};
                \node[circle, draw, fill = black, inner sep=1pt] at (9.5,0) (node c2) {}; \node at (9.5,-0.3) {\small $c$}; \node at (9.8,0.3) {\small $-r_{a}-r_s$};
                \node at (8.5,1) {$G_1$}; \draw[thin] (node a2) -- (node c2);
                \draw[thick] (10.5,0) -- (11.5,0);\draw[thick] (11,0.5) -- (11,-0.5);
                \node[circle, draw, fill = black, inner sep=1pt] at (12.5,0) (node c3) {}; \node at (12.5,-0.3) {\small $c$}; \node at (12.5,0.3) {\small $r_{a}+r_s$};
                \node[circle, draw, fill = black, inner sep=1pt] at (14.5,0) (node b3) {}; \node at (14.5,-0.3) {\small $b$}; \node at (14.5,0.3) {\small $r_{b}$};
                \node at (13.5,1) {$G_2$}; \draw[thin] (node c3) -- (node b3);
            \end{tikzpicture}
        \end{figure}
        \vspace{-5mm}
        $$DP(G',R_{G'}) = DP(G_1,(r_a,r_s,-r_a-r_s)) + DP(G_2, (r_a+r_s,r_b))$$ \\

\newpage
        \item \textbf{Case 2.3} $G'$ is formed using a parallel composition of $G_1$ and $G_2$.
        
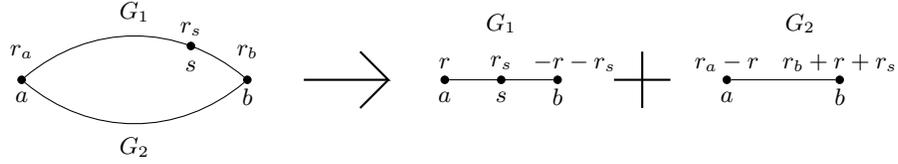
\begin{figure}[H]
            \centering
            \begin{tikzpicture}[scale = 0.75]
                \node[circle, draw, fill = black, inner sep=1pt] at (0,0) (node a) {}; \node at (0,-0.3) {\small $a$}; \node at (0,0.5) {\small $r_{a}$};
                \node[circle, draw, fill = black, inner sep=1pt] at (4,0) (node b) {}; \node at (4,-0.3) {\small $b$}; \node at (4,0.5) {\small $r_{b}$};
                \node[circle, draw, fill = black, inner sep=1pt] at (3,0.6) (node s) {}; \node at (3,0.25) {\small $s$}; \node at (3,0.9) {\small $r_{s}$};
                \draw[bend right = 40,thin] (node a) to (node b); \draw[bend left = 40,thin] (node a) to (node b);
                \node at (2,1.2) {$G_1$}; \node at (2,-1.2) {$G_2$};
                \draw[thick] (5,0) -- (6.5,0) -- (6,0.5);\draw[thick] (6.5,0) -- (6,-0.5);
                \node[circle, draw, fill = black, inner sep=1pt] at (7.5,0) (node a2) {}; \node at (7.5,-0.3) {\small $a$}; \node at (7.5,0.3) {\small $r$};
                \node[circle, draw, fill = black, inner sep=1pt] at (8.5,0) (node s2) {}; \node at (8.5,-0.3) {\small $s$}; \node at (8.5,0.3) {\small $r_s$};
                \node[circle, draw, fill = black, inner sep=1pt] at (9.5,0) (node c2) {}; \node at (9.5,-0.3) {\small $b$}; \node at (9.8,0.3) {\small $-r-r_s$};
                \node at (8.5,1) {$G_1$}; \draw[thin] (node a2) -- (node c2);
                \draw[thick] (10.5,0) -- (11.5,0);\draw[thick] (11,0.5) -- (11,-0.5);
                \node[circle, draw, fill = black, inner sep=1pt] at (12.5,0) (node c3) {}; \node at (12.5,-0.3) {\small $a$}; \node at (12.5,0.3) {\small $r_{a}-r$};
                \node[circle, draw, fill = black, inner sep=1pt] at (14.5,0) (node b3) {}; \node at (14.5,-0.3) {\small $b$}; \node at (14.5,0.3) {\small $r_{b} + r + r_s$};
                \node at (13.8,1) {$G_2$}; \draw[thin] (node c3) -- (node b3);
            \end{tikzpicture}
        \end{figure}
        \vspace{-5mm}
        $$DP(G',R_{G'}) = \min_{-F\leq r\leq F}DP(G_1,(r,r_s,-r-r_s)) + DP(G_2,(r_a-r,r_b+r+r_s))$$
    \end{itemize}
    \item \textbf{Case 3} $R_{G'}$ is a 4-tuple i.e $R_{G'}=(r_{a},r_s,r_t,r_{b})$ \\
    \begin{itemize}
        \item \textbf{Case 3.1} $G'$ is formed using a series composition of $G_1$ and $G_2$ splitting at $s$ (or $t$).
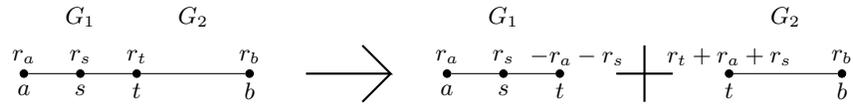
        \begin{figure}[H]
            \centering
            \begin{tikzpicture}[scale = 0.75]
                \node[circle, draw, fill = black, inner sep=1pt] at (0,0) (node a) {}; \node at (0,-0.3) {\small $a$}; \node at (0,0.3) {\small $r_{a}$};
                \node[circle, draw, fill = black, inner sep=1pt] at (1,0) (node s) {}; \node at (1,-0.3) {\small $s$}; \node at (1,0.3) {\small $r_{s}$};
                \node[circle, draw, fill = black, inner sep=1pt] at (2,0) (node c) {}; \node at (2,-0.3) {\small $t$}; \node at (2,0.3) {\small $r_{t}$};
                \node[circle, draw, fill = black, inner sep=1pt] at (4,0) (node b) {}; \node at (4,-0.3) {\small $b$}; \node at (4,0.3) {\small $r_{b}$};
                \draw[thin] (node a) -- (node c) -- (node b);
                \node at (1,1) {$G_1$}; \node at (3,1) {$G_2$};
                \draw[thick] (5,0) -- (6.5,0) -- (6,0.5);\draw[thick] (6.5,0) -- (6,-0.5);
                \node[circle, draw, fill = black, inner sep=1pt] at (7.5,0) (node a2) {}; \node at (7.5,-0.3) {\small $a$}; \node at (7.5,0.3) {\small $r_{a}$};
                \node[circle, draw, fill = black, inner sep=1pt] at (8.5,0) (node s2) {}; \node at (8.5,-0.3) {\small $s$}; \node at (8.5,0.3) {\small $r_{s}$};
                \node[circle, draw, fill = black, inner sep=1pt] at (9.5,0) (node c2) {}; \node at (9.5,-0.3) {\small $t$}; \node at (9.8,0.3) {\small $-r_{a}-r_s$};
                \node at (8.5,1) {$G_1$}; \draw[thin] (node a2) -- (node c2);
                \draw[thick] (10.5,0) -- (11.5,0);\draw[thick] (11,0.5) -- (11,-0.5);
                \node[circle, draw, fill = black, inner sep=1pt] at (12.5,0) (node c3) {}; \node at (12.5,-0.3) {\small $t$}; \node at (12.5,0.3) {\small $r_t+r_{a}+r_s$};
                \node[circle, draw, fill = black, inner sep=1pt] at (14.5,0) (node b3) {}; \node at (14.5,-0.3) {\small $b$}; \node at (14.5,0.3) {\small $r_{b}$};
                \node at (13.5,1) {$G_2$}; \draw[thin] (node c3) -- (node b3);
            \end{tikzpicture}
        \end{figure}
        \vspace{-5mm}
        $$DP(G',R_{G'}) = DP(G_1,(r_a,r_s,-r_a-r_s)) + DP(G_2,(r_t+r_a+r_s,r_b))$$ \\

        \item \textbf{Case 3.2} $G'$ is formed using a series composition of $G_1$ and $G_2$ splitting $s$ and $t$ into different subgraphs.
        
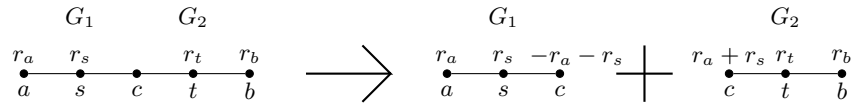
\begin{figure}[H]
            \centering
            \begin{tikzpicture}[scale = 0.75]
                \node[circle, draw, fill = black, inner sep=1pt] at (0,0) (node a) {}; \node at (0,-0.3) {\small $a$}; \node at (0,0.3) {\small $r_{a}$};
                \node[circle, draw, fill = black, inner sep=1pt] at (1,0) (node s) {}; \node at (1,-0.3) {\small $s$}; \node at (1,0.3) {\small $r_{s}$};
                \node[circle, draw, fill = black, inner sep=1pt] at (2,0) (node c) {}; \node at (2,-0.3) {\small $c$};
                \node[circle, draw, fill = black, inner sep=1pt] at (3,0) (node t) {}; \node at (3,-0.3) {\small $t$}; \node at (3,0.3) {\small $r_{t}$};
                \node[circle, draw, fill = black, inner sep=1pt] at (4,0) (node b) {}; \node at (4,-0.3) {\small $b$}; \node at (4,0.3) {\small $r_{b}$};
                \draw[thin] (node a) -- (node c) -- (node b);
                \node at (1,1) {$G_1$}; \node at (3,1) {$G_2$};
                \draw[thick] (5,0) -- (6.5,0) -- (6,0.5);\draw[thick] (6.5,0) -- (6,-0.5);
                \node[circle, draw, fill = black, inner sep=1pt] at (7.5,0) (node a2) {}; \node at (7.5,-0.3) {\small $a$}; \node at (7.5,0.3) {\small $r_{a}$};
                \node[circle, draw, fill = black, inner sep=1pt] at (8.5,0) (node s2) {}; \node at (8.5,-0.3) {\small $s$}; \node at (8.5,0.3) {\small $r_{s}$};
                \node[circle, draw, fill = black, inner sep=1pt] at (9.5,0) (node c2) {}; \node at (9.5,-0.3) {\small $c$}; \node at (9.8,0.3) {\small $-r_{a}-r_s$};
                \node at (8.5,1) {$G_1$}; \draw[thin] (node a2) -- (node c2);
                \draw[thick] (10.5,0) -- (11.5,0);\draw[thick] (11,0.5) -- (11,-0.5);
                \node[circle, draw, fill = black, inner sep=1pt] at (12.5,0) (node c3) {}; \node at (12.5,-0.3) {\small $c$}; \node at (12.5,0.3) {\small $r_{a}+r_s$};
                \node[circle, draw, fill = black, inner sep=1pt] at (13.5,0) (node t2) {}; \node at (13.5,-0.3) {\small $t$}; \node at (13.5,0.3) {\small $r_t$};
                \node[circle, draw, fill = black, inner sep=1pt] at (14.5,0) (node b3) {}; \node at (14.5,-0.3) {\small $b$}; \node at (14.5,0.3) {\small $r_{b}$};
                \node at (13.5,1) {$G_2$}; \draw[thin] (node c3) -- (node b3);
            \end{tikzpicture}
        \end{figure}
        \vspace{-5mm}
        $$DP(G',R_{G'}) = DP(G_1,(r_a,r_s,-r_a-r_s)) + DP(G_2,(r_a+r_s,r_t,r_b))$$ \\

\newpage

        \item \textbf{Case 3.3} $G'$ is formed using a series composition of $G_1$ and $G_2$ not splitting $s$ and $t$ into different subgraphs.
        
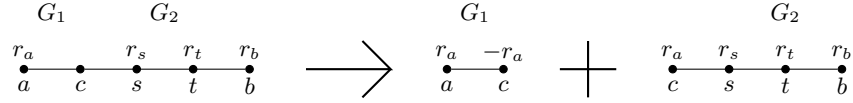
\begin{figure}[H]
            \centering
            \begin{tikzpicture}[scale = 0.75]
                \node[circle, draw, fill = black, inner sep=1pt] at (0,0) (node a) {}; \node at (0,-0.3) {\small $a$}; \node at (0,0.3) {\small $r_{a}$};
                \node[circle, draw, fill = black, inner sep=1pt] at (2,0) (node s) {}; \node at (2,-0.3) {\small $s$}; \node at (2,0.3) {\small $r_{s}$};
                \node[circle, draw, fill = black, inner sep=1pt] at (1,0) (node c) {}; \node at (1,-0.3) {\small $c$};
                \node[circle, draw, fill = black, inner sep=1pt] at (3,0) (node t) {}; \node at (3,-0.3) {\small $t$}; \node at (3,0.3) {\small $r_{t}$};
                \node[circle, draw, fill = black, inner sep=1pt] at (4,0) (node b) {}; \node at (4,-0.3) {\small $b$}; \node at (4,0.3) {\small $r_{b}$};
                \draw[thin] (node a) -- (node c) -- (node b);
                \node at (0.5,1) {$G_1$}; \node at (2.5,1) {$G_2$};
                \draw[thick] (5,0) -- (6.5,0) -- (6,0.5);\draw[thick] (6.5,0) -- (6,-0.5);
                \node[circle, draw, fill = black, inner sep=1pt] at (7.5,0) (node a2) {}; \node at (7.5,-0.3) {\small $a$}; \node at (7.5,0.3) {\small $r_{a}$};
                \node[circle, draw, fill = black, inner sep=1pt] at (8.5,0) (node c2) {}; \node at (8.5,-0.3) {\small $c$}; \node at (8.5,0.3) {\small $-r_{a}$};
                \node at (8,1) {$G_1$}; \draw[thin] (node a2) -- (node c2);
                \draw[thick] (9.5,0) -- (10.5,0);\draw[thick] (10,0.5) -- (10,-0.5);
                \node[circle, draw, fill = black, inner sep=1pt] at (11.5,0) (node c3) {}; \node at (11.5,-0.3) {\small $c$}; \node at (11.5,0.3) {\small $r_a$};
                \node[circle, draw, fill = black, inner sep=1pt] at (12.5,0) (node s3) {}; \node at (12.5,-0.3) {\small $s$}; \node at (12.5,0.3) {\small $r_s$};
                \node[circle, draw, fill = black, inner sep=1pt] at (13.5,0) (node t2) {}; \node at (13.5,-0.3) {\small $t$}; \node at (13.5,0.3) {\small $r_t$};
                \node[circle, draw, fill = black, inner sep=1pt] at (14.5,0) (node b3) {}; \node at (14.5,-0.3) {\small $b$}; \node at (14.5,0.3) {\small $r_{b}$};
                \node at (13.5,1) {$G_2$}; \draw[thin] (node c3) -- (node b3);
            \end{tikzpicture}
        \end{figure}
        \vspace{-5mm}
        $$DP(G',R_{G'}) = DP(G_1,(r_a,-r_a)) + DP(G_2,(r_a,r_s,r_t,r_b))$$ \\
        \item \textbf{Case 3.4} $G'$ is formed using a parallel composition of $G_1$ and $G_2$ with $s$ and $t$ in the same parallel component.
        
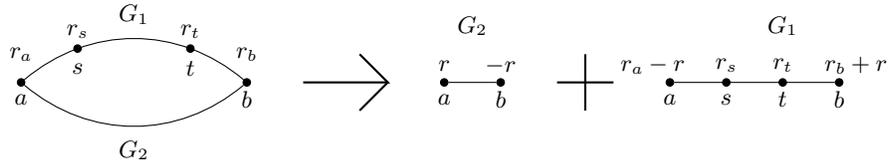
\begin{figure}[H]
            \centering
            \begin{tikzpicture}[scale = 0.75]
                \node[circle, draw, fill = black, inner sep=1pt] at (0,0) (node a) {}; \node at (0,-0.3) {\small $a$}; \node at (0,0.5) {\small $r_{a}$};
                \node[circle, draw, fill = black, inner sep=1pt] at (4,0) (node b) {}; \node at (4,-0.3) {\small $b$}; \node at (4,0.5) {\small $r_{b}$};
                \node[circle, draw, fill = black, inner sep=1pt] at (3,0.6) (node t) {}; \node at (3,0.25) {\small $t$}; \node at (3,0.9) {\small $r_{t}$};
                \node[circle, draw, fill = black, inner sep=1pt] at (1,0.6) (node s) {}; \node at (1,0.25) {\small $s$}; \node at (1,0.9) {\small $r_{s}$};
                \draw[bend right = 40,thin] (node a) to (node b); \draw[bend left = 40,thin] (node a) to (node b);
                \node at (2,1.2) {$G_1$}; \node at (2,-1.2) {$G_2$};
                \draw[thick] (5,0) -- (6.5,0) -- (6,0.5);\draw[thick] (6.5,0) -- (6,-0.5);
                \node[circle, draw, fill = black, inner sep=1pt] at (7.5,0) (node a2) {}; \node at (7.5,-0.3) {\small $a$}; \node at (7.5,0.3) {\small $r$};
                \node[circle, draw, fill = black, inner sep=1pt] at (8.5,0) (node c2) {}; \node at (8.5,-0.3) {\small $b$}; \node at (8.5,0.3) {\small $-r$};
                \node at (8,1) {$G_2$}; \draw[thin] (node a2) -- (node c2);
                \draw[thick] (9.5,0) -- (10.5,0);\draw[thick] (10,0.5) -- (10,-0.5);
                \node[circle, draw, fill = black, inner sep=1pt] at (11.5,0) (node c3) {}; \node at (11.5,-0.3) {\small $a$}; \node at (11.2,0.3) {\small $r_a-r$};
                \node[circle, draw, fill = black, inner sep=1pt] at (12.5,0) (node s3) {}; \node at (12.5,-0.3) {\small $s$}; \node at (12.5,0.3) {\small $r_s$};
                \node[circle, draw, fill = black, inner sep=1pt] at (13.5,0) (node t2) {}; \node at (13.5,-0.3) {\small $t$}; \node at (13.5,0.3) {\small $r_t$};
                \node[circle, draw, fill = black, inner sep=1pt] at (14.5,0) (node b3) {}; \node at (14.5,-0.3) {\small $b$}; \node at (14.8,0.3) {\small $r_{b}+r$};
                \node at (13.5,1) {$G_1$}; \draw[thin] (node c3) -- (node b3);
            \end{tikzpicture}
        \end{figure}
        \vspace{-5mm}
        $$DP(G',R_{G'}) = \min_{-F\leq r\leq F} DP(G_1,(r,-r)) + DP(G_2,(r_a-r,r_s,r_t,r_b+r))$$\\

        \item \textbf{Case 3.5} $G'$ is formed using a parallel composition of $G_1$ and $G_2$ with $s$ and $t$ in different parallel components.
        
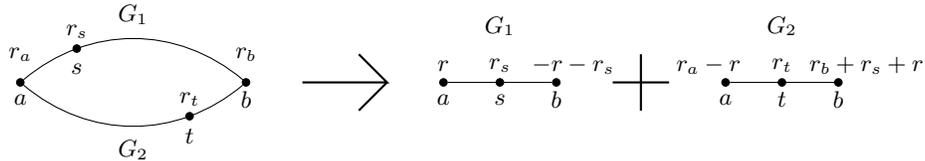
\begin{figure}[H]
            \centering
            \begin{tikzpicture}[scale = 0.75]
                \node[circle, draw, fill = black, inner sep=1pt] at (0,0) (node a) {}; \node at (0,-0.3) {\small $a$}; \node at (0,0.5) {\small $r_{a}$};
                \node[circle, draw, fill = black, inner sep=1pt] at (4,0) (node b) {}; \node at (4,-0.3) {\small $b$}; \node at (4,0.5) {\small $r_{b}$};
                \node[circle, draw, fill = black, inner sep=1pt] at (3,-0.6) (node t) {}; \node at (3,-0.95) {\small $t$}; \node at (3,-0.3) {\small $r_{t}$};
                \node[circle, draw, fill = black, inner sep=1pt] at (1,0.6) (node s) {}; \node at (1,0.25) {\small $s$}; \node at (1,0.9) {\small $r_{s}$};
                \draw[bend right = 40,thin] (node a) to (node b); \draw[bend left = 40,thin] (node a) to (node b);
                \node at (2,1.2) {$G_1$}; \node at (2,-1.2) {$G_2$};
                \draw[thick] (5,0) -- (6.5,0) -- (6,0.5);\draw[thick] (6.5,0) -- (6,-0.5);
                \node[circle, draw, fill = black, inner sep=1pt] at (7.5,0) (node a2) {}; \node at (7.5,-0.3) {\small $a$}; \node at (7.5,0.3) {\small $r$};
                \node[circle, draw, fill = black, inner sep=1pt] at (8.5,0) (node s2) {}; \node at (8.5,-0.3) {\small $s$}; \node at (8.5,0.3) {\small $r_{s}$};
                \node[circle, draw, fill = black, inner sep=1pt] at (9.5,0) (node c2) {}; \node at (9.5,-0.3) {\small $b$}; \node at (9.8,0.3) {\small $-r-r_s$};
                \node at (8.5,1) {$G_1$}; \draw[thin] (node a2) -- (node c2);
                \draw[thick] (10.5,0) -- (11.5,0);\draw[thick] (11,0.5) -- (11,-0.5);
                \node[circle, draw, fill = black, inner sep=1pt] at (12.5,0) (node c3) {}; \node at (12.5,-0.3) {\small $a$}; \node at (12.2,0.3) {\small $r_{a}-r$};
                \node[circle, draw, fill = black, inner sep=1pt] at (13.5,0) (node t2) {}; \node at (13.5,-0.3) {\small $t$}; \node at (13.5,0.3) {\small $r_t$};
                \node[circle, draw, fill = black, inner sep=1pt] at (14.5,0) (node b3) {}; \node at (14.5,-0.3) {\small $b$}; \node at (15,0.3) {\small $r_{b}+r_s+r$};
                \node at (13.5,1) {$G_2$}; \draw[thin] (node c3) -- (node b3);
            \end{tikzpicture}
        \end{figure}
        \vspace{-5mm}
        $$DP(G,R_{G'}) = \min_{-F\leq r\leq F}DP(G_1,(r,r_s,-r-r_s)) + DP(G_2,(r_a-r,r_t,r_b+r_s+r))$$
        \end{itemize}
\end{itemize}
We have exhausted all cases barring any symmetric ones and have shown how the DP table can be built. The total number of states in the DP are $O(mF^3)$ since the number of subgraphs $G'$ in the series-parallel decomposition tree is $O(m)$ \cite{valdes1979recognition} and the number of tuples $R_{G'}$ for a fixed subgraph $G'$ is bounded by $O(F^3)$ (by the condition that $\sum_{r\in R_{G'}}r = 0$). The above cases show that each state of the DP table can be calculated in $O(F)$ time. Hence, the total running time of computing the DP table is $O(mF^4)$. Finally, the optimal solution for the BCMFP will correspond to the largest value $r$ such that the cost of $DP(G,(0,r,-r,0))$ is bounded by the budget $B$. The optimal solution for the CapNDP will correspond to the value of $DP(G,(0,D,-D,0))$. This concludes the proof. \qed
\end{proof}

\section{Fully Polynomial Time Approximation Scheme}

In this section we prove Theorem \ref{thm:fptas} and provide an FPTAS for the BCMFP on series-parallel graphs by leveraging the pseudopolynomial time algorithm presented in the previous section.

\begin{proof}
Consider the following question,

\textit{Given a budget $C$ and a flow value $R$, is there a subset of edges $E'\subseteq E$ of cost at most $C$ such that there is a flow of value $R$ between $s$ and $t$ in $(V,E')$}?

The DP presented in the previous section will also be able to provide the subset of edges $E'$ if the answer to the above question is YES. Furthermore, the DP will run in time $O(mR^4)$ since we are only interested now in tuples $R_{G'}$ whose entries are bounded in absolute value by $R$.

Let OPT be the max flow possible within the budget $B$ and let $E^*\subseteq E$ be an optimal subset of the edges. Let $\epsilon>0$ be the desired level of accuracy and let $\epsilon'=\min(1,\epsilon/3)$. For any $M\in [1,F]$, define new capacities on the edges of $G$ as:
$$u_e^M:=\Bigl\lfloor{\frac{mu_e}{M\epsilon'}}\Bigr\rfloor$$

We now run the DP to answer the question above with capacities on edges given by $u_e^M$ and $R=\lceil m/\epsilon' \rceil$ and the budget $B$ remaining the same. As observed earlier, the running time of the DP will be $O(m^5/\epsilon'^4)$. Additionally, if $M\leq OPT/(1+\epsilon')$, then the answer to the above question will be YES. This is because the set of edges $E^*$ itself achieves a flow of value at least $R$. Indeed, let $(S,T)$ be an arbitrary cut in $(V,E^*)$ separating the source and the sink. Then, the capacity across this cut denoted $u^M(S,T)$ is given by,
\begin{align*}
    u^M(S,T)&=\sum_{\substack{(v,w)\in E^* \\ v\in S, w\in T}} \Bigl\lfloor{\frac{mu_{vw}}{M\epsilon'}}\Bigr\rfloor \\
    &\geq \frac{m}{M\epsilon'}u(S,T) - m \\
    &\geq \frac{m}{M\epsilon'}OPT - m \\
    &\geq \frac{m}{\epsilon'} && \text{(if $M\leq OPT/(1+\epsilon')$)}
\end{align*}

\noindent But since $u^M(S,T)$ is integral, we obtain $u^M(S,T)\geq \lceil m/\epsilon' \rceil = R$. Since the cut was arbitrary, we see that $(V,E^*)$ admits a flow of value at least $R$.

We can run a binary search in the search space $1,1+\epsilon',(1+\epsilon')^2,...$ up to $F$ to find the largest $M'$ such that we get YES as the answer to the question above when capacities are $u_e^{M'}$. From the arguments above, we know $M'\geq OPT/(1+\epsilon')^2$. Also let $E'$ be the corresponding optimal edge set found by running the DP (on the instance where $R=\lceil m/\epsilon' \rceil$ and capacities are $u_e^{M'}$). For each cut $(S,T)$ separating the source and the sink in $(V,E')$, we have,
\begin{align*}
    u(S,T)&=\sum_{\substack{(v,w)\in E' \\ v\in S, w\in T}}u_{vw} \\
    &\geq \frac{M'\epsilon'}{m}u^{M'}(S,T)\geq \frac{M'\epsilon'}{m}\frac{m}{\epsilon'} \geq \frac{OPT}{(1+\epsilon')^2} \\
    &\geq \frac{OPT}{(1+\epsilon)} && \text{since $\epsilon'=\min(1,\epsilon/3)$}
\end{align*}
The running time of the entire algorithm is polynomial in $m$, $1/\epsilon$ and $\log(F)$. since we are running the DP at most $\log(F)$ times and each run takes time $O(m^5/\epsilon^4)$. This concludes the proof. \qed
\end{proof}

\noindent \textbf{Remark:} The above method does not provide an FPTAS for the CapNDP on undirected series parallel graphs (in contrast to the results in \cite{krumke1999flow} for directed graphs). This is because we do not have the luxury of scaling the capacities in the CapNDP, as doing so would affect the feasible region of the problem instance. The objective function values can be scaled down and hence, only costs can be scaled in the CapNDP. In \cite{krumke1999flow}, the authors provided a different pseudopolynomial time algorithm with running time polynomial in $m$ and the maximum cost on an edge $c_{max}$. They could then scale down the costs to provide an FPTAS for CapNDP on directed series parallel graphs. It is currently unknown if the same can be done for undirected series-parallel graphs. In \cite{carr1999strengthening}, the authors provided a pseudopolynomial time algorithm for the CapNDP on outerplanar graphs and claimed that this can be used to obtain an FPTAS. However, this claim is erroneous since their pseudopolynomial time algorithm also had running time polynomial in $m$ and $F$. It is thus an interesting open question as to whether an FPTAS exists for CapNDP on undirected series parallel graphs or whether it is \textsc{APX-Hard}.

\section{Further Extensions}

\subsection{Capacities from a Lattice}

We argue that the pseudopolynomial time algorithm presented in Section 3 can be implemented in polynomial time under some assumptions on the capacities on edges. For instance, if the number of distinct values of capacities on edges is small, then intuitively it feels like the problem should be easier. We substantiate this by proving Theorem \ref{thm:lattice}. Let the capacities on the edges lie on a lattice defined by some non-negative numbers acting as a basis $d_1,\ldots, d_k$, i.e. every capacity $u_e$ is expressible as $u_e = \alpha_1d_1 + \cdots +\alpha_kd_k$ where $|\alpha_i|\leq K$ for all $i$. We assume that the number of basis elements $k$ is a constant, and the value of $K$ is polynomially bounded. We show that the algorithm provided in Section 3 can be implemented in polynomial time and hence, both the BCMFP and CapNDP are polynomial time solvable.

\begin{proof}
    The DP has states $(G',R_{G'})$ where $G'$ is a subgraph in the series-parallel decomposition tree of the input graph $G$, and $R_{G'}$ defines the residues in a circulation on $G'$. These residues are computed for all integer values from $-F$ to $F$ where $F$ is the max $s$-$t$ flow in the graph $G$. We argue that not every integer value from $-F$ to $F$ need to be considered. Indeed for any demanded flow $R$, let $E^*$ be an optimal solution to the CapNDP and consider the maximum $s$-$t$ flow in the graph $H=(V,E^*)$. Consider now a flow decomposition of this maximum flow. Each $s$-$t$ path in this flow decomposition saturates an edge of the graph $H$. Hence, the flow value on any edge in the graph $H$ lies in the set
    \[
    \{\sum_{i=1}^k\alpha_id_i\;:\;|\alpha_i|\leq mK\}
    \]
    This implies that in any subgraph $G'$ arising in the series-parallel decomposition tree of $G$, the residues on any node can be expressed as $\sum_{i=1}^k\alpha_id_i$ where each $|\alpha_i|$ is bounded by $m^2K$. The number of such residues is at most $O((m^2K)^k)$. Thus instead of considering $O(F^3)$ many values for $R_{G'}$, we only need to consider polynomially many. This completes the proof. \qed
\end{proof}

\subsection{Edge Upgrades}

We consider a variant of the BCMFP and CapNDP problems where each edge has some upgrade options wherein a cost can be paid to increase the capacity of the edge by a certain amount. However, at most one upgrade can be applied. Note that adding parallel edges does not solve this variant since then we cannot control how many upgrades are applied. We provide a gadget reduction from this variant to the original BCMFP and CapNDP. The gadget we provide is a series-parallel graph and hence, all our earlier results in this paper extend to this new variant of BCMFP and CapNDP on undirected series parallel graphs.

Formally, for every edge $e$ in $G$, there are various choices of costs and capacities provided as input $(c^1_e,u^1_e), \ldots, (c^k_e,u^k_e)$, and at most one of these choices can be purchased by a solution. We depict below the gadget construction to capture this variant when there are 2 or three choices, and this easily generalizes to more choices. We assume that $u^i\geq u^j$ if $i\geq j$.

\begin{figure}[H]
    \centering
\begin{tikzpicture}[scale=1, transform shape]
  \tikzset{every node/.style={circle, draw, minimum size=2mm, inner sep=0pt}}
  \node (n1) at (0,0) {};
  \node (n2) at (2,0) {};
  \node (n3) at (4,0) {};
  \node (n4) at (6,0) {};

  \path[every node/.style={font=\sffamily\small}]
    (n1) edge node[above] {0, $u_2$} (n2)
    (n2) edge[bend left] node[above] {$c_1$, $u_1$} (n3)
    (n2) edge[bend right] node[below] {$c_2$, $u_2$} (n3)
    (n3) edge node[above] {0, $u_2$} (n4);
\end{tikzpicture}
    \caption{An edge upgrade gadget with two choices}
\end{figure}
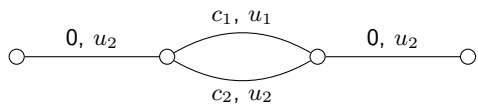

\begin{figure}
    \centering
\begin{tikzpicture}[scale=1, transform shape]
  \tikzset{every node/.style={circle, draw, minimum size=2mm, inner sep=0pt}}
  \node (n0) at (-2,0) {};
  \node (n1) at (0,0) {};
  \node (n2) at (2,0) {};
  \node (n3) at (4,0) {};
  \node (n4) at (6,0) {};
  \node (n5) at (8,0) {};

  \path[every node/.style={font=\sffamily\small}]
    (n0) edge node[above] {0, $u_3$} (n1)
    (n1) edge[bend right] node[below] {$c_3$, $u_3$} (n4)
    (n1) edge node[above] {0, $u_2$} (n2)
    (n2) edge[bend left] node[above] {$c_1$, $u_1$} (n3)
    (n2) edge[bend right] node[below] {$c_2$, $u_2$} (n3)
    (n3) edge node[above] {0, $u_2$} (n4)
    (n4) edge node[above] {0, $u_3$} (n5);
\end{tikzpicture}
\caption{An edge upgrade gadget with three choices}
\end{figure}
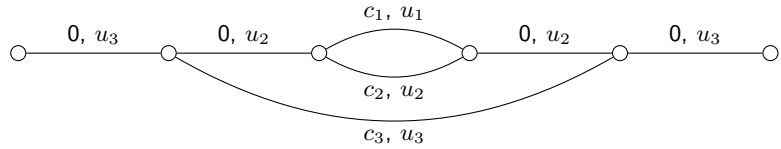

Note that, for $k$ upgrade choices of an edge, the above reduction creates $k + 2(k-1)$ edges and $2k$ nodes, thus increasing the size of the input graph linearly. Furthermore, we are replacing an edge of the graph with a series-parallel graph. Hence, if the original graph was series-parallel, then after applying our reduction the graph will remain series-parallel.

\bibliographystyle{splncs04}
\bibliography{ISCO}

\begin{thebibliography}{1}
\providecommand{\url}[1]{\texttt{#1}}
\providecommand{\urlprefix}{URL }
\providecommand{\doi}[1]{https://doi.org/#1}

\bibitem{carr1999strengthening}
Carr, R.D., Fleischer, L.K., Leung, V.J., Phillips, C.A.: Strengthening
  integrality gaps for capacitated network design and covering problems. Tech.
  rep., Sandia National Lab.(SNL-NM), Albuquerque, NM (United States);
  Sandia~… (1999)

\bibitem{chakrabarty2015approximability}
Chakrabarty, D., Chekuri, C., Khanna, S., Korula, N.: Approximability of
  capacitated network design. Algorithmica  \textbf{72},  493--514 (2015)

\bibitem{holzhauser2016budget}
Holzhauser, M., Krumke, S.O., Thielen, C.: Budget-constrained minimum cost
  flows. Journal of Combinatorial Optimization  \textbf{31}(4),  1720--1745
  (2016)

\bibitem{holzhauser2017network}
Holzhauser, M., Krumke, S.O., Thielen, C.: A network simplex method for the
  budget-constrained minimum cost flow problem. European journal of operational
  research  \textbf{259}(3),  864--872 (2017)

\bibitem{holzhauser2017complexity}
Holzhauser, M., Krumke, S.O., Thielen, C.: On the complexity and
  approximability of budget-constrained minimum cost flows. Information
  Processing Letters  \textbf{126},  24--29 (2017)

\bibitem{krumke1999flow}
Krumke, S.O., Noltemeier, H., Schwarz, S., Wirth, H.C., Ravi, R.: Flow
  improvement and network flows with fixed costs. In: Operations Research
  Proceedings 1998: Selected Papers of the International Conference on
  Operations Research Zurich, August 31--September 3, 1998. pp. 158--167.
  Springer (1999)

\bibitem{schwarz1998budget}
Schwarz, S., Krumke, S.O.: On budget-constrained flow improvement. Information
  Processing Letters  \textbf{66}(6),  291--297 (1998)

\bibitem{valdes1979recognition}
Valdes, J., Tarjan, R.E., Lawler, E.L.: The recognition of series parallel
  digraphs. In: Proceedings of the eleventh annual ACM symposium on Theory of
  computing. pp. 1--12 (1979)

\end{thebibliography}

\end{document}